\documentclass[english]{lni}
\usepackage[anythingbreaks]{breakurl}
\begin{document}
\title[When Audits and Recounts are a Distraction]{When Audits and Recounts Distract from Election Integrity:
The 2020 U.S.\ Presidential Election in Georgia
  }
\subtitle{To appear in Proceedings of E-Vote-ID 2024, LNI} 
 \author[1]{Philip B. Stark}{pbstark@berkeley.edu}{0000-0002-3771-9604}
 \affil[1]{Department of Statistics\\University of California\\Berkeley, CA 94720-3860\\USA}
\maketitle

\begin{abstract}
The U.S.\ state of Georgia was central to efforts to overturn the results of the 2020 Presidential election,
including a phone call from then-president Donald Trump to Georgia Secretary of State
Brad Raffensperger asking Raffensperger to `find' 11,780 votes.
Raffensperger has maintained that a `100\% full-count risk-limiting audit'
and a machine recount agreed with the initial machine-count results, which proved that the reported election results were accurate
and that `no votes were flipped.' 
While there is no evidence that the reported outcome is wrong, neither is there evidence that it is correct:
the two machine counts and the manual `audit' tallies disagree substantially, even about the number of ballots cast.
Some ballots in Fulton County, Georgia, were included in the original count at least twice; some were included in the machine recount at least thrice.
Audit handcount results for some tally batches were omitted from the reported audit totals: reported audit results do not include all the 
votes the auditors counted.
In short, the two machine counts and the audit were not probative of who won because of poor processes and controls: 
a lack of secure physical chain of custody, ballot accounting, pollbook reconciliation, and accounting for other election materials such as memory cards. 
Moreover, most voters used demonstrably untrustworthy ballot-marking devices; as a result, even a perfect handcount or audit
would not necessarily reveal who really won. 
True risk-limiting audits (RLAs) and rigorous recounts can limit the risk that an incorrect electoral outcome will be certified rather than being corrected. 
But no procedure can limit that risk without
a trustworthy record of the vote.
And even a properly conducted RLA of some contests in an election does not show that any other contests in that election
were decided correctly.
 The 2020 U.S.\ Presidential election in Georgia illustrates unrecoverable errors that can render
 recounts and audits `security theater' that distract from the more serious problems rather than justifying trust.
\end{abstract}


\section{Introduction: The 2020 U.S.\ Presidential Election in Georgia}
Georgia was one of the `swing states' that determined the outcome of the 2020 U.S.\ presidential election:
its 16 electoral college votes went to Joe Biden.
In a well publicized recording of then-president Donald Trump to Georgia Secretary of State Brad Raffensperger,
Trump asked Raffensperger to `find' 11,700 votes.\footnote{%
See, e.g., \url{https://int.nyt.com/data/documenttools/highlights-of-trump-s-call-with-the-georgia-secretary-of-state-1/b67c0d9dbde1a697/full.pdf}
visited 11~July 2024.
Subsequently, in early 2021, Trump-affiliated parties gained improper access to all components of
the voting system in Coffee County, Georgia and copied and distributed the codebase and data.
See, e.g., \url{https://www.cnn.com/2023/08/13/politics/coffee-county-georgia-voting-system-breach-trump/index.html},
\url{https://apnews.com/article/2022-midterm-elections-technology-lawsuits-donald-trump-voting-6a1324cc6cf45c95ca086a5c81617b15},
\url{https://www.washingtonpost.com/investigations/2022/10/28/coffee-county-election-voting-machines/}, all accessed 11~July 2024.
}

Georgia performed a second machine count and hired VotingWorks to orchestrate a `risk-limiting audit' of the 2020 presidential contest, including providing software.
This paper shows that the audit does not support the election results; that the election, recount, and audit disagree;
and that all three were unreliable.
Among other issues,
  some memory cards containing votes were not uploaded in the first machine count.
Some ballots were included in the first machine tally at least twice.
  Some ballots were included in the second machine tally at least three times.
   And some votes manually tabulated in the audit
were not included in the reported audit totals.
  Moreover, the state of Georgia requires all in-person voters to use Dominion Voting Systems (DVS) ballot marking devices (BMDs) to mark their ballots.
These devices are vulnerable to hacking and misconfiguration \cite{haldermanReport23};
voters rarely check BMD printout \cite{bernhardEtal20,kortumEtal20,haynesHood21}; when voters do check, they are unlikely to notice and report printing errors \cite{bernhardEtal20,kortumEtal20};
and no feasible amount of pre-election testing, logic and accuracy testing, or election-day monitoring can suffice to show that BMDs misbehavior did not alter the outcome \cite{starkXie22}.
BMD printout is thus not a trustworthy basis for evidence-based elections \cite{starkWagner12,appelEtal20,appelStark20}, even when voted ballots are
curated adequately and proper procedures are followed.
While there is no evidence of widespread fraud, the mismanagement of the election, reliance on untrustworthy vote records, 
lack of physical controls on ballots and other voting materials, lack of sanity checks, and poorly executed procedures make it 
impossible to know who ``really'' won.

This story is about Georgia, but the moral is broader:
some of the things that can and do go wrong in administering elections result in an untrustworthy vote record.
Auditing a poorly run election with an untrustworthy vote record is a distraction from the fact that the vote record is not trustworthy,
not a way to justify trust.
Auditing cannot restore trustworthiness to a poorly run election; rather, it is a way to ``tie a bow around'' a \emph{well-run} election
to show that whatever might have gone wrong did not alter the electoral outcome.

\section{The 2020 audit}
Secretary of State Brad Raffensperger claimed, ``Georgia's historic first statewide audit reaffirmed that
  the state's new secure paper ballot voting system accurately counted
  and reported results. \textellipsis
 [W]e did a 100 percent
risk-limiting audit with a hand recount which proved the accuracy of
the count and also proved that the machines were accurately counting
it, and that no votes were flipped.''\footnote{\url{https://sos.ga.gov/news/historic-first-statewide-audit-paper-ballots-upholds-result-presidential-race},
accessed 11~July 2024.} 
  VotingWorks Executive Director Ben
  Adida claimed ``Georgia's first statewide audit successfully confirmed
  the winner of the chosen contest and should give voters increased
  confidence in the results.''\footnote{Ibid.} 
  Per the official report
  of the audit, ``[t]he audit confirmed the original result of the
  election, namely that Joe Biden won the Presidential Contest in the
  State of Georgia. 
  The audit {[}{]} provides sufficient evidence that
  the correct winner was reported.''\footnote{Ibid.} 

Secretary Raffensperger has also used the recount and audit in his defense against a lawsuit that seeks to
provide all Georgia voters the option to hand-mark paper ballots in person, rather than being compelled to use BMDs 
(Curling et al. v. Raffensperger et al.,  Civil Action No. 1:17-CV-2989-AT,
U.S.\ District Court for the Northern District of Georgia, Atlanta Division).
Raffensperger has publicly painted the opposing election security experts in this matter---some of the world's top cybersecurity experts---as ``stop-the-steal'' conspiracy theorists, muddying the waters with false claims about the recount and audit and 
deliberately conflating ``there is strong evidence that the election was poorly run and little evidence that the outcome is correct'' 
with ``there is strong evidence that the outcome is wrong and that fraud was committed.''
Some of the data analyzed below (cast vote records, in particular) were obtained in discovery in \emph{Curling v.\ Raffensperger}, but most are a matter of public record
and can be downloaded from the Georgia Secretary of State's website, from URLs given below.

The so-called `risk-limiting audit' did not limit the risk of certifying an incorrect electoral outcome for many reasons,
starting with its reliance on an untrustworthy record of the votes.
The record is untrustworthy because of how it was created (largely BMD printout), curated (a lack of physical accounting
for ballots and other materials, lack of pollbook reconciliation, and other elements of a proper canvass),
and organized (no ``ballot manifest'').
The audit \emph{could} have checked the tabulation of the validly
cast ballots it found, but it did not check that properly, as proved by documents on the Secretary
of State's website.\footnote{%
    \url{https://sos.ga.gov/news/historic-first-statewide-audit-paper-ballots-upholds-result-presidential-race}
    accessed 11~July 2024.
    Audit data at the urls  \url{https://sos.ga.gov/admin/uploads/Georgia\%202020\%20RLA\%20Report.xlsx},
    \url{https://sos.ga.gov/admin/uploads/county-summary-data.pdf}, and
    \url{https://sos.ga.gov/admin/uploads/audit-report-November-3-2020-General-Election-2020-11-19.csv},
    linked from that page, are periodically unavailable, producing the message ``Sorry, you have been blocked.
    You are unable to access sos.ga.gov.''
    RLA manual tabulation batch
    sheets were downloaded from    \url{https://sos.ga.gov/admin/uploads/Fulton\%20RLA\%20Batches.zip} on
    9~January 2022.
    Precinct-level results for the original machine tally 
    are at \url{https://results.enr.clarityelections.com//GA//105369/271927/reports/detailxls.zip}; 
    for the second machine tally, they are at \url{https://results.enr.clarityelections.com//GA//107231/273078/reports/detailxls.zip},
    both visited 2~September 2024.
}

\subsection{Things the audit did not check}
The audit did not check whether BMDs correctly printed voters' selections. 
No audit can check that \cite{appelEtal20}.
(As a consequence, Secretary Raffensperger had no basis to assert that
no votes were flipped.)
Expert declarations and testimony in \emph{Curling v.\ Raffensperger} establish that 
the Dominion BMDs can be hacked, misprogrammed, or
misconfigured to print votes that differ from voters' selections as
confirmed onscreen or through audio. 
Logic and accuracy testing cannot establish that BMDs
behave correctly in practice \cite{starkXie22}.  
Only voters are in a
position to check---but few do, and those who do check generally check
poorly (see citations below).
Georgia has no
procedures to log, investigate, or report complaints from
voters that BMDs altered votes, so it is unknown whether voters
observed problems. 

\begin{itemize}
\item The audit did not check whether every validly cast ballot was included in the tally
exactly once. 
The audit \emph{could not} check whether every validly cast
ballot was scanned, because Georgia's rules for ballot accounting,
pollbook and voter participation reconciliation, physical chain of
custody, etc., do not account for every cast ballot.

\item The audit did not check whether the number of participating voters differed from the number
of cast ballots.
  
\item The audit did not check whether every memory card used in the election
was accounted for, nor whether every memory card containing votes was
uploaded to a tabulator. 
During the audit, it was discovered that some cards had not
been uploaded, but there was no comprehensive check to confirm that
every card was eventually included exactly once.
Below are examples of ballots that were erroneously included in machine counts more than once.

\item The audit did not check whether scans were duplicated, deleted,
replaced or altered.

\item  The audit did not check whether QR-encoded votes on BMD
printout match the human-readable selections on any ballot.

\item The audit did not check whether the voting system correctly
interpreted any ballot or BMD printout.

\item The audit did not aggregate its own manual tallies correctly, as explained below.

\end{itemize}

\noindent
The analysis below focuses on Fulton County (Atlanta), but 
there is no reason to believe the problems are confined to Fulton; indeed,
lapses such as failing to upload memory cards occurred in other counties.
  
\subsection{The audit report omitted some batch tallies}
The audit was conducted using ``sort and stack'':
teams sorted batches of  ballots (including BMD printout) by the presidential vote, then counted the
sorted stacks.
Batch tallies were manually entered on paper  `Audit Board Batch Sheets,'  (ABBSs). 
Other workers transcribed the ABBSs
into VotingWorks audit software ``Arlo'' to create a database of tallies; totals were calculated from that database.
A spreadsheet of results was produced from Arlo.
Every ballot validly cast in Fulton County in the 2020 Presidential
Election should be reflected in exactly one ABBS, and data from every
ABBS should have been (but was not) entered exactly once into the database from
which the audit spreadsheet was exported.
The transcription of ABBSs was not observable by the public, but the public could in principle compare
posted images of the ABBSs to the posted audit spreadsheet, as described below.
(Spoiler alert: they do not match.)

Many ABBSs were not completely filled in. 
The ``Batch Type,''
signifying the mode of voting (absentee, election day, advance) was
often blank, as were many places numbers belonged.
The four posted ABBS image files for Fulton County contain a total of 1,927 ABBSs.\footnote{%
    Audit subtotals come from the detailed ``audit spreadsheet'' available at
    \url{https://sos.ga.gov/admin/uploads/audit-report-November-3-2020-General-Election-2020-11-19.csv}
    accessed 11~July 2024.
    Images of the Fulton County, GA, RLA manual tabulation batch
    sheets (ABBSs) were downloaded from \url{https://sos.ga.gov/admin/uploads/Fulton\%20RLA\%20Batches.zip} on
    9~January 2022. 
    That file contains five .pdf files, ``Fulton Audit
    Documents 1\_redacted.pdf,'' through ``Fulton Audit Documents
    4\_redacted.pdf,'' which contain images of ABBSs, and ``Fulton Audit
    Documents 5.pdf'' which contains images of ``Vote Review Panel Tally
    Sheets.''
} 
But the audit spreadsheet contains only 1,916 rows of data for Fulton County. 
At least eleven ABBSs are entirely missing, not counting
possible duplicate entries in the spreadsheet.\footnote{%
    However, there is at least one ABBS marked ``Dup'' (presumably meaning
    ``duplicate'') for instance, page 11 of ``Fulton Audit Documents
    2\_redacted.pdf.'' However, as table~\ref{tab:missing-abbss} shows, at least 11~ABBSs are not
    accounted for in the audit spreadsheet. 
    Thus, there are presumably
    duplicated entries in the audit spreadsheet.
} 
This sort of ``sanity
check'' is simple to perform, but apparently was not performed by the
auditors, VotingWorks, Fulton County, or the Secretary of State.

Table~\ref{tab:missing-abbss} lists 11~ABBSs that do not appear in the audit spreadsheet; the final column indicates
which page of which ABBS image file contains the image (for instance,
``4 at 162'' means page 162 of ``Fulton Audit Documents
4\_redacted''). 
The scans of the ABBSs are available at \url{https://figshare.com/s/9819e969a8a6172c25bc} (Appendix~1).
The fact that the vote data in the last two rows are
identical is suspicious, but the corresponding ABBS images are clearly
different.
Regardless, neither appears in the audit spreadsheet.

\begin{table}
\centering
\tiny{
\begin{tabular}{cccccccccccc}
& Scanner &  Batch & Mode of voting & Trump & Biden & Jorgensen & Write-In & Undervote or blank & Overvote & Image source \\
\hline
1 & 3 & 48 & absentee & 4 & 93 & 2 & 0 & 0 & 0 & 4 at 162
\\
2
 & 2 & 52 & absentee
 & 
6
 & 92
 & 0
 & 0
 & 0
 & 0
 & 1 at 1
\\
3
 & 
3
 & 
12--14
 & 
?
 & 
12
 & 
83
 & 
1
 & 
0
 & 
0 & 0 & 
4 at 128 \\
4
 & 
3
 & 
239
 & 
?
 & 
13
 & 
87
 & 
0
 & 
0
 & 
0
 & 
0
 & 
3 at 177 \\
5
 & 
1
 & 
80--84
 & ?  & 118 & 329 & 3 & 2 & 2
 & 
1
 & 
3 at 519 \\
6
 & 
3
 & 
260
 & 
absentee
 & 
30
 & 
66
 & 
0
 & 
0
 & 
0
 & 
0
 & 
4 at 355  \\
7
 & 
 & 
AP01A-1
 & 
election day
 & 
84
 & 
62
 & 
6
 & 
2
 & 
1
 & 
0
 & 
1 at 170  \\
8
 & 
3
 & 
179--181
 & 
absentee
 & 
85
 & 
224
 & 
5
 & 
1
 & 
2
 & 
0
 & 
4 at 293  \\
9
 & 
2
 & 
239
 & 
absentee
 & 
4
 & 
42
 & 
0
 & 
0
 & 
0
 & 
0
 & 
2 at 153  \\
10
 & 
Chastain
 & 
12
 & 
advance
 & 
613
 & 
605
 & 
24
 & 
7
 & 
4
 & 
0
 & 
3 at 351 \\
11
 & 
Chastain
 & 
114
 & 
advance
 & 
613
 & 
605
 & 
24
 & 
?
 & 
4
 & 
0
 & 
3 at 270
\end{tabular}
}
\caption{ \protect \label{tab:missing-abbss} Examples of audit board batch sheets (ABBSs, tallies of votes in batches of ballots) that were not entered into the audit results spreadsheet.}
\end{table}

There are no data in the audit spreadsheet
matching rows 4--11 of table~\ref{tab:missing-abbss}.
There are data that match rows 1, 2,
and 3, but with different batch identifiers.\footnote{%
   The data that match row~1 are identified as ``Scanner 3 Ballot {[}sic{]}
    162'' rather than batch~48. 
    The data that match row~2 are identified
    as ``Absentee Scanner 2 Batch 400'' rather than batch 52. 
    The data
    that match row~3 are identified as Absentee Scanner 3 Batch 253
    rather than batches 12--14.
} 
There is no reason to doubt that these are
genuinely different batches: some identical counts in different batches are to be
expected. 
Indeed, in the entire audit spreadsheet, there are 16,807
rows that duplicate other ABBS vote counts within the same county, out
of a total of 41,881 rows.
  
Vote totals for Trump, Biden, and
Jorgensen derived by summing ABBS entries in the audit spreadsheet
match the vote totals in the summary audit result spreadsheet posted
by the Secretary of State at the URL
\url{https://sos.ga.gov/admin/uploads/Georgia\%202020\%20RLA\%20Report.xlsx},
downloaded on 9~January 2022. 
The spreadsheet does not list
write-ins, undervotes, or overvotes.
Both sources show Trump receiving
137,620 votes, Biden receiving 381,179, and Jorgensen receiving 6,494.
Thus, the ABBSs that are missing from the audit spreadsheet are also
missing from the audit's reported vote totals.
  
On the assumption that the ABBSs---the original source of the manual
tally data entered into the audit spreadsheet---are correct, the
omission of that sample of 11 ABBSs deprived Trump of 1,582 votes,
Biden of 2,288, and Jorgensen of 65, not to mention write-ins. 
This
sample alone has a total of over 3,900 votes that the audit tabulated
but were not included in Fulton County's audit vote totals, compared with
a \emph{statewide} margin of less than 12,000 votes.
  
The original tabulation in Fulton County showed 524,659 votes; the
reported audit results showed 525,293, a difference of 634 votes,
about 0.12 percent.\footnote{%
    Data from
\url{https://sos.ga.gov/admin/uploads/Georgia\%202020\%20RLA\%20Report.xlsx},
    accessed 9~January 2022.
} 
Accounting for those 11~omitted ABBSs
increases the apparent tabulation error from 634 votes to over
4,569 votes or 0.87 percent, far larger than the statewide margin of
victory. 
It is also larger than 0.73 percent, which Secretary
Raffensperger claimed was the maximum miscount in any Georgia
county.\footnote{%
    Per Secretary Raffensperger, ``{[}i{]}n Georgia's
    recount, the highest error rate in any county recount was 0.73\%.''
\url{https://sos.ga.gov/index.php/elections/2020\_general\_election\_risk-limiting\_audit},
    accessed 9~January 2022.
}
  
There is no way to know whether including those 11
ABBSs would make the audit tabulation a complete count in
Fulton County: many
ballots might remain untabulated, because Georgia's canvass procedures are lax.
The proof some
Georgia jurisdictions do not keep adequate track of ballots,
memory cards, and other election materials is evidenced by the fact
that thousands of ballots and scans were ``discovered'' during the
audit.\footnote{%
\url{https://www.cbs46.com/news/floyd-county-election-director-fired-after-audit-reveals-2-600-votes-went-uncounted/article\_bbd08d90-2aa2-11eb-9e4d-bf96ac56ad54.html},
    accessed 10~January 2022.
\url{https://www.news4jax.com/news/georgia/2020/11/18/4th-georgia-county-finds-uncounted-votes-as-hand-count-deadline-approaches/},
    accessed 10~January 2022.
\url{https://www.mdjonline.com/elections/cobb-elections-finds-350-uncounted-ballots-during-audit/article\_0d93e26e-22bd-11eb-8bce-17067aceee33.html},
    accessed 10~January 2022.
\url{https://www.11alive.com/article/news/politics/elections/fayette-county-election-results-ballots-uncovered-during-audit/85-f79dd838-a15c-4407-80b2-9dfbc2466188},
    accessed 10~January 2022.
}
There is no trustworthy inventory of
ballots to check the results against.
  
Georgia Governor Brian P.\ Kemp pointed out similar flaws in the audit,
saying the audit report was ``sloppy, inconsistent, and presents
questions about what processes were used by Fulton County to arrive at
the result.''\footnote{%
    Letter from Brian P. Kemp, Governor, to the
    Georgia State Election Board, dated 17 November 2021, addressing the
    work of Mr. Joseph Rossi; Review of Inconsistencies in the Data
    Supporting the Risk Limiting Audit Report, Office of Governor Brian
    P. Kemp, 17~November 2021.
} 
Governor Kemp's letter points out that the audit data
include duplicated entries.

\section{First Count, Audit, and Recount Differ Substantially}
Official
  precinct-level results for the original tabulation were downloaded from
  \url{https://results.enr.clarityelections.com//GA/Fulton/105430/271723/reports/detailxls.zip}
  and for the recount from
  \url{https://results.enr.clarityelections.com//GA/Fulton/107292/275183/reports/detailxls.zip}
  to examine the results in precinct RW01, the precinct in which the lead plaintiff in \emph{Curling v.\ Raffensperger} votes.

  Table~\ref{tab:rwo1} shows the counts of election-day votes in 
  precinct RW01 for the three presidential candidates, according
  to the original machine count, the machine recount, and the ``audit,''
  and vote-by-mail and advance votes for the original election and the
  recount. 
  (The audit did not report precinct-level results for
  vote-by-mail or advance voting.)

\begin{table}
\tiny{
\begin{tabular}{c|ccc|ccc|ccc|ccc}
Count & \multicolumn{3}{c}{Election Day} & \multicolumn{3}{c}{Advance} & \multicolumn{3}{c}{Absentee by Mail} & \multicolumn{3}{c}{Provisional} \\
\hline
& Trump & Biden & Jorgensen & Trump & Biden & Jorgensen & Trump & Biden
& Jorgensen & Trump & Biden & Jorgensen \\
\hline
Original & 193 & 88 & 11 & 1455 & 1003 & 23 & 619 & 833 & 15 & 9 & 4 &
1 \\
Recount & 162 & 73 & 9 & 1487 & 1015 & 25 & 619 & 809 & 15 & 5 & 3 &
1 \\
Audit & 243 & 88 & 11 & & & & & & & & & \\
\end{tabular}
}
\caption{\protect \label{tab:rwo1} Election day, advance, absentee, and provisional vote tallies for Fulton County, GA, precinct RW01 in the 2020 U.S.\ Presidential election}
\end{table}

There are large, unexplained differences among these
results.\footnote{%
    There appears to be some cancellation of error, but
    the hand count kept ballots cast in different ways separated
    (advance in-person, absentee by mail, and election day). It
    is not clear how misclassification of the mode of voting would
    affect one candidate's totals much more than the other candidates.
    Regardless, these discrepancies are large and should be
    investigated, including inspecting the physical ballots.
  } 
Secretary Raffensperger attributed all differences between the audit
and the original count to human counting error, citing a 2012 study
that found hand-count error rates as high 
as 2~percent.\footnote{%
\url{https://sos.ga.gov/index.php/elections/historic\_first\_statewide\_audit\_of\_paper\_ballots\_upholds\_result\_of\_presidential\_race},
    accessed 10~January 2022.
} 
While human error presumably accounts for \emph{some} of the
  difference, there is no evidence that it accounts for most of the
  difference, much less the entire difference, as Secretary
  Raffensperger claimed.\footnote{%
  Moreover, RLAs treat the hand count as the correct count: the hand counts should be conducted with
  adequate care to ensure they are accurate, which typically requires different procedures
  from those used in initial manual tallies.
}
  
The original count and audit agree with each other (but not with the
  recount) regarding the number of election-day votes for Biden and
  Jorgensen. 
  The audit found 50 more election-day votes for Trump than
  the original tally and 81 more than the machine recount found: a
  difference of almost 50 percent. 
  These differences have not been
  investigated and are unexplained. 
  A hypothesized error rate of 2
  percent in hand counts does not suffice.
  
  The differences might result from
  discrepancies between the QR-encoded votes and the human-readable
  votes on BMD printout and/or from misconfiguration, bugs, or malware
  on the scanners or tabulators. 
  As discussed above, the audit checked
  none of these things. 
  Possible machine error should have been investigated, rather than assumed not to exist.
  
The hand count could easily be more accurate than the machine count.
Indeed, it is well known that careful hand counts of hand-marked paper ballots
are often more accurate than machine counts, in part because human
readers can interpret faint, improper, and ambiguous marks better than
machines can, even when the machines are working properly,
as studies of ``residual votes'' and statewide recounts show
\cite{ansolabehereReeves04,ansolabehereEtal18,ansolabehereStewart05,alvarezEtal13a,alvarezEtal13b,carrier05}.\footnote{%
    Whether hand counts are more accurate than machine counts depends on many variables. The scrutiny and care involved in recounts and manual audits are generally higher than they are in initial hand counts.
    For instance, \cite{ansolabehereEtal18} find that \emph{initial} machine counts were often more accurate than \emph{initial} hand counts---by using careful handcounts from statewide recounts as the touchstone for the correct counts.
}

The scanner settings Georgia uses for its Dominion scanners 
(low resolution, black-and-white) can cause voters' selections not to
appear at all in the images, selections that are obvious to human readers
looking at the actual ballots.\footnote{%
    See, e.g.,
    Judge Amy Totenberg's Opinion and Order of 11~October 2020, in
    Curling et al.\ v.\ Raffensperger, 1:17-CV-2989-AT,  at 4, 30, 95, 101, 103, 114--135.
  }
Manual tallies generally find more
  valid votes than machine tallies. 
  Hand-count error rates are known to depend on many factors, including
  ballot design, the method for hand counting (``sort-and-stack'' versus
  ``read-and-mark''), and the size of counting teams. 
  They
  presumably also depend on whether there are additional quality control
  measures in place, such as checking sorted piles of ballots to ensure
  that each pile has votes for just one candidate.
  
  The study \cite{gogginEtal12} cited by
  Secretary Raffensperger to support his claim is a laboratory study with 108 subjects and 120
  ballots, each containing 27 contests with two candidates. 
  It used
  three kinds of ``ballots'': printout from two kinds of DRE
  (direct-recording electronic) voting system and an optical scan
  ballot. 
  The highest error rates were for thermal printout from DREs,
  which does not resemble Georgia's BMD printout nor Georgia's
  hand-marked paper ballots.
   The method with the \emph{highest} error rate was the
  ``sort-and-stack'' tally method that Georgia used in its
  audit. 
  The study did not observe hand tabulation in a real
  election, nor did it involve BMD summary printout. 
  
  The differences between the original count and the machine recount are
  large and unexplained; for instance, the difference  in the
  counts of Biden's Absentee votes is almost 3 percent. 
  It is now impossible to
  know what went wrong, nor whether the differences are primarily
  attributable to malware, bugs, misconfiguration, or human error.

\subsection{The two machine counts in Fulton County}
  
  This section assesses the internal consistency of the two machine counts (the
  original machine count and the machine recount) in Fulton County
  using data from the  election management
  system (EMS) including cast vote records (CVRs),
  scanned images of ballots, and BMD printout, and other files made available to
  the plaintiffs in Curling et al.\ v.\ Raffensperger et al.
  To confirm that the EMS data were the correct data, tallies were calculated and compared to the official results
  for Fulton County; they matched:\footnote{First machine count results:
    \url{https://results.enr.clarityelections.com/GA/Fulton/105430/web.264614/\#/summary}
    (visited 11~July 2024) Second machine count results:
    \url{https://results.enr.clarityelections.com/GA/107231/web.264614/\#/detail/5000?county=Fulton}
    (visited 11~July 2024)
}

\begin{table}
\centering
\begin{tabular}{crr}
Candidate & \multicolumn{1}{c}{1st machine count} & \multicolumn{1}{c}{2nd machine count} \\
\hline
Donald J. Trump & 137,240 & 137,247 \\
Joseph R. Biden & 381,144 & 380,212 \\
Jo Jorgensen & 6,275 & 6,320 \\
\end{tabular}
\caption{\protect \label{tab:machine-totals} Data used to verify that the EMS download matched the official results in Fulton County, GA.}
\end{table}
  
  The number of cast vote records (the voting system's record of the
  votes on each ballot or BMD printout card, from which the system
  tabulates results) in the two machine counts in Fulton County were
  rather different: 528,776 in the first count and 527,925 in the second
  count, a difference of 851. 
  Fulton County has not explained this discrepancy.
  
  The number of cast vote records in the two machine counts should be
  equal. 
  Differences might occur if (i)~some ballots or BMD printout
  cards were misplaced or found between the two machine counts, so a
  different number pieces of paper was scanned in the two machine
  counts; (ii)~malware, bugs, misconfiguration, or a bad actor added,
  deleted, or altered records in the election management system in one
  or both machine counts; (iii)~Fulton County did not scan every validly
  cast ballot or BMD printout card exactly once in each machine count; (iv) some scans were omitted or improperly included in one or both counts.
  Compelling evidence that (ii) or (iii) is true is presented below, but
  all four could be true simultaneously.
    
  Fulton County did not produce an image file for every
  cast vote record. 
  For the first machine count, production included
  images of ballots or BMD printout cards for only 168,726 of the
  528,776 cast vote records: 376,863 image files are missing. 
  For the
  second machine count, Fulton County's production included images of
  ballots or BMD printout cards for 510,073 of the 527,925 cast vote
  records: 17,852 image files are missing.
  
  Entire batches of images are missing from Fulton County's production;
  for example, images from Scanner 801 batch 117 and Scanner 801 batch
  118 are referred to in the cast vote records for the second machine
  count but the images were not among the electronic records. 
  Without
  additional information it is impossible to determine whether the missing
  images are missing because of human error or malfeasance, programming errors (bugs),
  or malware in Fulton County's election management system (EMS)---possibilities that
 are not mutually exclusive.
  
  The extant images nonetheless prove that
  Fulton County's election results included many votes more than once in
  the reported tabulations. 
  The full extent of multiple-counting cannot be determined without additional information, but there
  is evidence that it added thousands of bogus votes to the
  reported machine-count results. 
  That is, thousands of Fulton County
  voters' votes were included in the reported totals more than once.
  It is not possible to determine
  conclusively whether any voter's votes were omitted from the reported
  totals.
    
  Repeatedly scanning the same piece of paper generally does not produce
  images that are bitwise identical, because of variations in the
  alignment of the paper, illumination within the scanner, dirt on
  scanner lenses, etc. Similarly, a single scan can be altered digitally
  to produce multiple images that look similar but are not bitwise
  identical.
  
  Small variations in voters' marks (e.g., not filling an oval
  completely or straying outside the oval) on hand-marked paper ballots
  generally make it possible to tell whether two separate scans of
  hand-marked paper ballots that contain the same votes are scans of the
  same physical ballot.
  
  It is not generally possible to tell whether two 200dpi
  black-and-white scans of BMD printout cards are scans of the same
  piece of paper simply by looking at those two scans, because BMD
  printout cards containing the same votes may be indistinguishable at low
  resolution in black-and-white.\footnote{%
  Differences in the monochrome threshold or scanner maintenance might create discernable differences. 
  A sufficiently high-resolution
    scan might make it possible to identify differences in the
    arrangement of the paper fibers \cite{clarksonEtal09}.
} 
   However, if both scans contain a rare write-in
  name or rare combination of write-in names, that is evidence of a
  duplicate. 
  Similarly, if a series of votes is repeated in in the same
  order (or reverse order) in different scan batches of BMD printout,
  that is also evidence that they are repeated images of the same
  collection of paper. If the duplicated (or reversed) vote sequences
  are long and include rare write-in names, the evidence that they are
  scans of the same physical pieces of paper is compelling.
  
  There are at least 12 hand-marked ballots from Fulton County
  precinct RW01 that were scanned twice in the first machine count (the
  original election). 
  Fourteen pairs of duplicate images are listed in table~\ref{tab:count-doubles}
  and are available at the url \url{https://figshare.com/s/9819e969a8a6172c25bc} (Appendix~2).
  The format of the numbers is $<$scanner number$>$\_$<$batch number$>$\_$<$image number$>$.
  At least three BMD cards from precinct
  RW01 appear to have been scanned twice in the machine
  recount in RW01, based on the votes and the order in which they were
  scanned in two batches. 
  In particular, Scanner 801, batches 43 and
  44---both comprising scans of advance in-person BMD printout
  cards---start with images of 214~BMD cards that have the same sets of votes in the same order in both batches. 
  The two batches were scanned within about five minutes of each other,
  according to the timestamps in the images. 
  Many of the images show
  write-in votes\footnote{%
  Write-ins included votes for ``Anyone,''
    ``XXX,'' ``Willie Nelson,'' and ``Alexander Hamilton,'' as well as
    write-in votes for ``Donald Trump'' for District Attorney, Clerk of
    the Superior Court, Tax Commissioner, Sheriff, Solicitor General,
    and Surveyor.
} 
or votes for third-party candidates, further evidence
  that the match was not coincidence. 
  Visual inspection of all 214~pairs and confirmed that they match: 
  those BMD cards were scanned twice in the machine
  recount. 
  The other 211 (214--3=211) duplicated scans are of BMD cards
  from other precincts in Fulton County.

\begin{table}
\centering
\begin{tabular}{c|c|c|}
pair & Image A & Image B \\
\hline
1 & 05162\_00234\_000096 & 05162\_00235\_000057 \\
2 & 05162\_00234\_000093 & 05162\_00235\_000054 \\
3 & 05162\_00234\_000074 & 05162\_00235\_000036 \\
4 & 05162\_00234\_000072 & 05162\_00235\_000034 \\
5 & 05162\_00234\_000068 & 05162\_00235\_000030 \\
6 & 05162\_00234\_000069 & 05162\_00235\_000031 \\
7 & 05162\_00234\_000054 & 05162\_00235\_000014 \\
8 & 05162\_00234\_000031 & 05162\_00235\_000090 \\
9 & 05162\_00234\_000026 & 05162\_00235\_000085 \\
10 & 05162\_00234\_000017 & 05162\_00235\_000076 \\
11 & 05162\_00234\_000013 & 05162\_00235\_000072 \\
12 & 05162\_00234\_000014 & 05162\_00235\_000073 \\
13 & 05162\_00234\_000003 & 05162\_00235\_000062 \\
14 & 05162\_00234\_000001 & 05162\_00235\_000060 \\
\end{tabular}
\caption{\protect \label{tab:count-doubles}Images that were included in the original machine count in Fulton County at least twice.
Images are posted at \url{https://figshare.com/s/9819e969a8a6172c25bc} (Appendix~2).}
\end{table}
    
  There is also one hand-marked paper ballot that
  was scanned twice in RW01 in the machine recount, and at least seven
  hand-marked paper ballots that were scanned thrice in RW01 in the
  machine recount. 
  Twenty-nine images seem to represent only 11
  distinct pieces of paper, even though they contributed 29~votes to some contests, including the presidential contest. 
  The sets of images are available at the url \url{https://figshare.com/s/9819e969a8a6172c25bc} (Appendix 3).  
  Table \ref{tab:recount_multiple} lists the pairs and triples.

\begin{table}
\centering
\begin{tabular}{l|c|c|c}
Multiple & Image A & Image B & Image C \\
\hline
1 & 00801\_00044\_000168 & 00801\_00043\_000168 & \\
2 & 00801\_00044\_000083 & 00801\_00043\_000083 & \\
3 & 00801\_00044\_000042 & 00801\_00043\_000042 & \\
4 & 05160\_00074\_000023 & 05160\_00067\_000008 & \\
5 & 00794\_00017\_000024 & 00791\_00026\_000091 &
00791\_00019\_000010 \\
6 & 00794\_00017\_000029 & 00791\_00026\_000086 &
00791\_00019\_000015 \\
7 & 00794\_00018\_000001 & 00791\_00026\_000009 &
00791\_00019\_000092 \\
8 & 00794\_00018\_000011 & 00791\_00026\_000019 &
00791\_00019\_000082 \\
9 & 00794\_00019\_000002 & 00791\_00026\_000079 &
00791\_00019\_000022 \\
10 & 00794\_00019\_000005 & 00791\_00026\_000076 &
00791\_00019\_000025 \\
11 & 00794\_00019\_000006 & 00791\_00026\_000075 &
00791\_00019\_000026 \\
\end{tabular}
\caption{\protect \label{tab:recount_multiple} Images that were (erroneously) included in the machine recount at least three times.
Images are posted at \url{https://figshare.com/s/9819e969a8a6172c25bc} (Appendix 3).}
\end{table}
  
  To confirm that the duplicate and triplicate images were included in
  the reported vote tabulation, the cast-vote records (CVRs)
  produced by Fulton County for each image identifier among the
  duplicates and triplicates of images of RW01 ballots and BMD printout
  cards were searched electronically. 
  All 24 from the original count and all 29  from the machine recount were among the CVRs. 
  Therefore,  the
  duplicate and triplicate votes were included in the reported machine
  tabulations, since the vote totals derived from the CVRs agree with
  the reported vote totals, as mentioned above.
  
  For Fulton County as a whole, plaintiffs 
  in \emph{Curling v.\ Raffensperger} 
  identified images of 2,871 ballots and BMD printout cards that they
  claim were counted two or three times in the second machine count.
  Some were identified by visual inspection of the images; others were
  inferred to be duplicates because a sequence of cast vote records was
  identical (or reversed) for long portions of two scan batches. 
  I confirmed that 214 of the
  purported duplicate scans of BMD cards were indeed duplicates. 
  This list of 2,871 is a sample from a larger list of
  images of ballots and BMD printout cards that plaintiffs
  assert were included in the tabulation twice or more.
  All 6,118 images in question were referenced in CVRs in
  the second machine count, so all contributed to the
  tabulation.
  
  Nine hundred sixteen (916) of the 2,871 sets of images were images of hand-marked paper ballots. 
  In a random sample of 100
  of those 916, I verified visually that 46 contained triplicate
  images.
 I confirmed the determination for 98 of the
  100 sets. 
 I disagreed about one  set, and
  was unable to verify one set. 
  Treating this conservatively as 98
  agreements in 100 random checks
 yields a 95 percent lower confidence
  bound that at least 891 of the 916 claimed multiples are genuine
  multiples.
  
  These observations make it clear that
  in the original
  count and in the machine recount, Fulton County did not keep track of which ballots and
  BMD cards had been scanned and which had not.  
  It is also possible that the
  electronic records were altered accidentally or intentionally, or that some memory cards were not uploaded or uploaded more than once. 
  The electronic records of the election are not intact. 
  This is a
  surprising gap: the
  most basic election safeguard is to check whether the number of voters
  who participated is equal to the number of ballots and BMD printout
  cards that were cast and to the number that were tabulated. 
  Moreover,
  one might reasonably expect all electronic election materials to be backed up
  onsite and offsite, at least for the U.S.\ federally mandated retention
  period of twenty-two months, so the loss of hundreds of thousands of
  image files from the first machine count and of nearly 18,000 images
  from the second machine count is hard to fathom.
  
  Fulton County would have noticed these errors if it had kept
  track of ballots and BMD printout cards and checked the total number
  against the number reported in the electronic tabulation. 
  It seems
  that Fulton County did not know how many ballots and BMD printout
  cards were cast in the election, how many voters cast votes, or how
  many pieces of paper were scanned---nor how those numbers compare to
  each other. 
  Absent basic ballot accounting, pollbook reconciliation,
  and counting of electronic records, it is unsurprising that the two
  machine tallies differ so much.
  The U.S.\ Election Assistance Commission has published
  best practices for chain of custody.\footnote{%
\url{https://www.eac.gov/sites/default/files/bestpractices/Chain\_of\_Custody\_Best\_Practices.pdf}
    accessed 11~July 2024.
}
  
  Fulton County's lax curation and processing of cast
  ballots, BMD printout, and electronic records make a true
  risk-limiting audit impossible because even a perfect tabulation of the votes from the available paper might not
  show who really won.
  Voters have good reason to believe that some votes counted more than
  others, since some votes were included twice or thrice in the totals. 
  There
  is no way to know how many votes were omitted from the tabulation,
  absent access to the physical ballots and BMD printout and evidence
  that the chain of custody is intact.
  It is impossible to determine whether malware, bugs, misconfiguration,
  or malfeasance disenfranchised voters or altered the election results.
  
  The audit planning, process, and controls did not detect
  the double and triple counting. 
  Even if Fulton County
  did not rely on ballot-marking devices for all in-person
  voters, the lack of basic accounting controls makes it impossible to
  determine who really won, even by a perfect hand count of the votes: the record of the vote could easily be incomplete or
  adulterated. 
  There is no reason to believe that problems of the kinds described above
  are limited to Fulton County.

\section{Summary}
  
  An accurate recount of the votes in a trustworthy record can determine the true winners of an election,
  and a rigorous audit can provide confidence that a well-run election found
  the true winner(s).
  But neither a recount nor an audit can compensate for using untrustworthy
  technology to record votes, for instance, because the election was run poorly and had inadequate physical security controls; in such
  circumstances, recounts and audits distract attention from the real problems rather
  than justifying confidence.
  Absent a trustworthy record of the votes, no
  procedure can provide affirmative evidence that the reported winner(s)
  really won. 
  
  Georgia lacks such a record for many reasons, including
  the heavy reliance on BMDs; lack of physical accounting of voted
  ballots, memory cards, and other election materials; lack of pollbook
  and voter participation reconciliation; lack of rigorous chain of custody; etc.  
  To provide reasonable assurance that every validly cast vote is
  counted---accurately---requires systematic improvements:

\begin{enumerate}
\item   
  Every voter should have the opportunity to mark a ballot by hand,
  whether voting in person in advance, in person on election day, or
  absentee by mail.
  
 \item  Reduce the use of ballot-marking devices to a minimum:\footnote{%
 Hand-marked ballots should be offered to in-person voters by default, with access to a BMD available upon request.
 BMDs or other accessible means of marking a ballot should be set up in advance, so that it is available if and when a voter requests to use it.
 BMD printout should resemble hand-marked paper ballots to the extent possible,
 to preserve voter privacy: 
 they should  the same paper stock, have the same format as hand-marked paper ballots, and the marks should be printed to resemble
 hand-made marks, e.g., by digitizing actual hand-made marks.
}
 \begin{itemize}
 \item  BMDs do not necessarily print voters' selections accurately. They
  can be hacked or misconfigured \cite{haldermanReport23,appelEtal20}.

  \item A growing body of empirical work shows that few voters check the
  BMD printout, and those who do rarely catch and report errors \cite{bernhardEtal20,kortumEtal20,haynesHood21}.

  \item There is no way for a voter to prove to an election official or
  anyone else that a BMD malfunctioned. Hence, there is no way to ensure that malfunctioning devices are removed
  from service if  voters notice BMDs misbehaving. And if
  a device is caught misbehaving, there is no way to reconstruct the
  correct election outcome \cite{appelEtal20}.

  \item There is no way to test BMDs adequately prior to, during, or after
  an election to establish whether they altered votes, even if they
  altered enough votes to change electoral outcomes \cite{appelEtal20,starkXie22}.

\end{itemize}

   \item  Implement better procedures and checks on chain of
  custody of election materials, especially voted ballots. 
  Georgia currently cannot  determine whether every validly cast
  ballot was included in the reported results
  exactly once, whether there was
  electronic or physical ``ballot-box stuffing,'' or whether votes were
  altered.\footnote{%
  This is evidenced by the fact that the 2020 audit
    found thousands of untabulated ballots. 
    Per the Secretary of State's office, ``{[}t{]}he audit process also
    led to counties catching making mistakes they made in their original
    count by not uploading all memory cards.''
   \url{https://sos.ga.gov/news/historic-first-statewide-audit-paper-ballots-upholds-result-presidential-race}
    accessed 11~July 2024.
    Because
    physical accounting for election materials was lacking, there is no way to know
    how many more votes validly cast in that election were not
    included in any of the reported tallies. 
    Moreover, the lax
    recordkeeping evidently resulted in scanning the same batches of
    ballots more than once. 
    Similarly, some ABBSs were presumably
    entered more than once, and as shown above, some were not entered at
    all.
} 
 
 \item  Implement better protocols for using and checking physical
  security seals on ballots and voting equipment---and check whether
  those protocols were followed. 
  Require routine
  scrutiny of custody logs and surveillance video, and
  other related security measures.
  
  \item Perform internal consistency checks 
  as part of the canvass, including, e.g.:

\begin{enumerate}
\item  
  Verify that the number of ballots sent to each polling location
  (and blank paper stock for ballot-marking devices and ballot-on-demand
  printers) equals the number returned voted, spoiled, or unvoted. 
  This check should be physical, based on manual inventories, not on reports from the voting system.
\item  
  Check pollbooks and other voter participation records against the
  number of voted ballots received, including whether the
  appropriate number of ballots of each ``style'' were received.
\item  
  Check whether the number of electronic vote records (images and CVRs) agrees with the physical inventory of ballots of
  each style.
\end{enumerate}
\end{enumerate}
  
A genuine risk-limiting audit requires a demonstrably trustworthy record of voter intent.
  Georgia's vote records are untrustworthy for many reasons, starting
  with the heavy use of ballot-marking devices, which do not produce a trustworthy record of the vote
  \cite{appelEtal20,haldermanReport23}
  no matter how much logic and accuracy testing or 
  election-day monitoring there is \cite{starkXie22}.
  The lack of a trustworthy record is exacerbated in Georgia by the lack of ballot accounting, pollbook reconciliation, and other elements of a good canvass.
  There
  are also problems with Georgia's verification of voter eligibility and voter participation record. 
  But even if every voter used a hand-marked paper
  ballot and there were no issues determining voter eligibility, Georgia
  does not keep track of election materials adequately
  through physical inventories, custody logs, and other means. 
  
  The foundation for a
  risk-limiting audit is a \emph{ballot manifest}, a physical inventory
  of the validly cast paper ballots detailing how they are stored: the number of containers, their identifiers, and the number of cards
  in each.   
  It must be derived without reliance on the voting system or the
  audit is trusting the voting system to check itself. 
  For example, if some cards were never scanned or some scans were not
  uploaded (as discovered during the 2020 ``audit''), they will be
  missing from any manifest derived from the voting system. 
  Absent a physical inventory, it is impossible to account for votes reliably and impossible
  to limit the risk that an incorrect electoral outcome will be
  certified, even with a careful manual recount or rigorous audit: recounting or applying risk-limiting audit procedures to an untrustworthy
  collection of ballots is ``security theater.'' 
  
Like many states, Georgia audits only a small number of contests in each election.
Even a properly conducted RLA using a demonstrably trustworthy paper trail confirms only the contest or contests
that were audited---and no other contests---although election officials sometimes claim otherwise.\footnote{%
  For example, the State of Colorado 
  currently conducts an RLA of two contests in each jurisdiction in each election, but the Secretary of State's website says,
  ``Colorado residents can be confident that official election results reflect the will of voters because we conduct a statewide bi-partisan audit after every election to ensure the integrity of the results.'' \url{https://www.coloradosos.gov/pubs/elections/auditCenter.html} 
  accessed 24~July 2024.
  }
  
  \emph{Acknowledgments.} This work was supported in part by NSF Grant SaTC–2228884. 

\bibliography{pbsbib}
\end{document}